\documentclass[twocolumn,showpacs,preprintnumbers,amsmath,amssymb,superscriptaddress]{revtex4}
\usepackage{graphicx}
\usepackage{dcolumn}
\usepackage{bm}
\usepackage{graphicx}
\usepackage{xspace}
\usepackage{multirow}
\usepackage{natbib}
\usepackage{color}

\begin{document}
\title{Unconventional superconducting gap in NaFe$_{0.95}$Co$_{0.05}$As observed by ARPES}

\author{Z.-H.~Liu} 
\affiliation{ Department of Physics, Renmin University, Beijing, 100872, China.}

\author{P.~Richard}
\affiliation{WPI Research Center, Advanced Institute for Material Research, Tohoku University, Sendai 980-8577, Japan.}
\affiliation{Beijing National Laboratory for Condensed Matter Physics, and Institute of Physics, Chinese Academy of Sciences Beijing 100190, China}

\author{K.~Nakayama}
\affiliation{Department of Physics, Tohoku University, Sendai 980-8578, Japan.}

\author{G.-F.~Chen}
\author{S.~Dong}
\author{J.-B.~He}
\author{D.-M.~Wang}
\author{T.-L.~Xia} 
\affiliation{ Department of Physics, Renmin University, Beijing, 100872, China.}
 
\author{K.~Umezawa}
\author{T.~Kawahara}
\affiliation{Department of Physics, Tohoku University, Sendai 980-8578, Japan.}

\author{S.~Souma}
\affiliation{WPI Research Center, Advanced Institute for Material Research, Tohoku University, Sendai 980-8577, Japan.}
\author{T.~Sato}
\affiliation{Department of Physics, Tohoku University, Sendai 980-8578, Japan.}
\affiliation{TRiP, Japan Science and Technology Agency (JST), Kawaguchi 332-0012, Japan.}
\author{T.~Takahashi}
\affiliation{WPI Research Center, Advanced Institute for Material Research, Tohoku University, Sendai 980-8577, Japan.}
\affiliation{Department of Physics, Tohoku University, Sendai 980-8578, Japan.}

\author{T. Qian}
\author{Yaobo~Huang}
\author{Nan~Xu}
\author{Yingbo~Shi}
\author{H.~Ding}
\affiliation{Beijing National Laboratory for Condensed Matter Physics, and Institute of Physics, Chinese Academy of Sciences Beijing 100190, China}

\author{S.-C. Wang}
\email{scw@ruc.edu.cn}
\affiliation{ Department of Physics, Renmin University, Beijing, 100872, China.}

\begin{abstract}
    We have performed high resolution angle-resolved photoemission measurements on superconducting electron-doped NaFe$_{0.95}$Co$_{0.05}$As ($T_{c}\sim$18 K). We observed  a hole-like Fermi surface around the zone center and two electron-like Fermi surfaces around the M point which can be connected by the $Q=(\pi, \pi)$ wavevector, suggesting that scattering over the near-nested Fermi surfaces is important to the superconductivity of this ``111" pnicitide. Nearly isotropic superconducting gaps with sharp coherent peaks are observed below $T_c$ on all three Fermi surfaces. Upon increasing temperature through $T_c$, the gap size shows little change while the coherence vanishes. Large ratios of $2\Delta/k_{B}T_{c}\sim8$ are observed for all the bands, indicating a strong coupling in this system. These results are not expected from a classical phonon-mediated pairing mechanism.
\end{abstract}
\maketitle

The superconducting (SC) energy gap is one of the most important quantities in revealling the pairing mechanism of  superconductors. For example, the
conventional BCS superconductors have isotropic $s$-wave SC gaps ($\Delta$) with a $2\Delta/k_BT_c \sim3.5$ ratio, a
characteristic of phonon-mediated pairing in the weak-coupling regime.
In contrast, the high-$T_{c}$ cuprate superconductors have an anisotropic $d$-wave gap with a much larger $2\Delta_0/k_B T_c$ ratio. 
The discovery of a new class of high-$T_{c}$ superconductors in Fe-pnictides \cite{Kamihara2008p1087} with many unconventional properties raises the question of novel pairing mechanism in these materials. Due to the complexity of these multi-band systems, no consensus has been reached yet on the SC gap symmetry and the pairing mechanism. In fact, literature contains many contradicting theoretical \cite{Chubukov2008lp,Raghu2008p4502,Daghofer2008p4503} and experimental results 
\cite{Szabo2009kv, Hashimoto2010p5612, Nakai2010p4517, Hicks2009p5418, Reid2010p64501, Malone2009p5413}. 
  
Angle-resolved photoemission spectroscopy (ARPES) is a powerful tool in probing directly the low-energy states of superconductors in the momentum ($k$)
space. It thus allows a direct band-selective measurement of the size and $k$-profile of the SC gap.
ARPES measurements on 122 \cite{Ding2008p3325,LiuHaiYun2008p3303}, 11 \cite{Nakayama2010p5613} and
1111 \cite{Kondo2008p3964} systems indicate isotropic nodeless gaps in the strong coupling regime ($2\Delta/k_{B}T_{c}\sim 5-7$).
Among theoretical models promoted to explain these results, antiferromagnetic (AF) fluctuations have been regarded as a major candidate for pairing.
Compared to other families of Fe-based superconductors, the 111 family has weaker magnetic correlations \cite{Li2009p2147}.
Recently, the consistency of the ARPES measurements on various families of Fe-based superconductors has been challenged by an ARPES report on the
LiFeAs system (111) \cite{Borisenko2010p4526}.
Phonon-mediated BCS pairing was proposed based on the observation of a $2\Delta/k_{B}T_{c}\sim3.1$ ratio \cite{Inosov2010p4498,Borisenko2010p4526}. 
   
Here we report ultra-high resolution ARPES measurements on electron-doped 111 pnicitide NaFe$_{0.95}$Co$_{0.05}$As ($T_{c}\sim18$ K).
One hole-like Fermi surface (FS) centered at the zone center ($\Gamma$) is quasi-nested to two elliptical FSs centered at the M point ($\pi$,~$\pi$).
Signatures of an additional small hole pocket at $\Gamma$ are also observed. The FSs have nearly isotropic gaps with variations smaller than $\pm$0.5 meV.
These results are consistent with a $s_\pm$~symmetry \cite{Sknepnek2009p3832,Kondo2008p3964} and exclude
the possibility of symmetries with nodes.
All gaps on hole and electron FSs have sizes of $\sim$6-7 meV, leading to a $2\Delta/k_{B}T_{c} \sim$8 ratio. The
temperature dependence of the SC quasiparticles is similar to that of underdoped high-$T_c$ cuprates.
Our results point towards unconventional superconductivity in the 111 family of Fe-based superconductors.

High quality single crystals of NaFe$_{1-x}$Co$_{x}$As (x$\approx$0.05) were synthesized by the flux method \cite{Chen2008rp}. The cleaved surface, which is expected to happen between the two weakly-bonding Na layers, is symmetrical and non-polarized surface which is usually bulk representative and has less surface disorder. This is reflected by clear band dispersion and sharp quasiparticle peak observed by ARPES in this material, as described below.
Samples were mounted in a gas protected glove box to prevent reaction with moisture.
Samples with size of 1$\times$1 mm$^2$ were cleaved \emph{in situ}, and yield flat, mirror like (001) surfaces.
ARPES measurements were done at Tohoku University used a Scienta 2002 analyzer with
a helium discharging lamp (He-I$\alpha$ line, $h\nu$ = 21.2 eV),
and at the Synchrotron Radiation Center, WI, using a Scienta R4000 analyzer. We performed experiments within the 4 --100 K temperature range and in a vacuum better than $4\times10^{-11}$~$Torr$.
No noticeable aging was observed during each measurement cycle. The energy resolution using for measuring fine features was better than 3 meV.

\begin{figure}[thb]
\begin{centering}
\includegraphics[width=3.4 in]{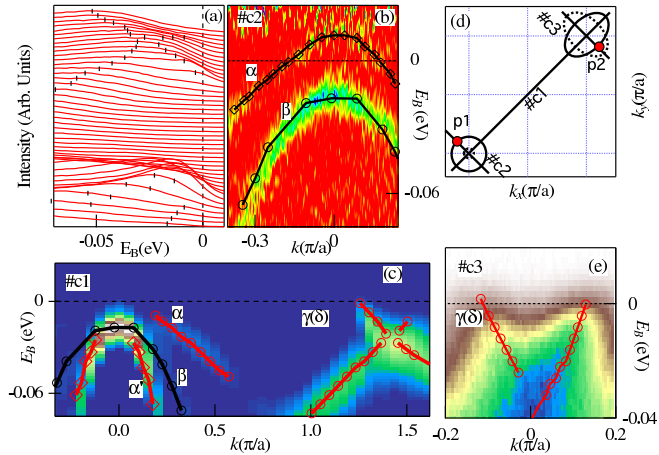} 
\end{centering}
\vspace{-2em}
\caption{
  (color online)
  (a) EDCs of NaFe$_{0.95}$Co$_{0.05}$As along (0, 0)--($\pi$, $\pi$) in the normal state ($T$ = 30 K), indicated as \#c1 in (d).
  (b) Second derivative ($\partial^2I(\omega, k)/\partial \omega^2$) of intensity along cut \#c2 in (d) at $T$ = 100 K. The Fermi
  function was removed to enhance the band dispersion above $E_F$. Markers give extracted band dispersions. 
  (c) Second derivative ($\partial^2I(\omega, k)/\partial \omega^2$) of intensity along along (0, 0)--($\pi$, $\pi$) in the normal state $T$ = 30 K.
  (d) Sketch of the FSs in NaFe$_{0.95}$Co$_{0.05}$As.
 (e) Photoemission intensity along cut \#c3 in (d) ($T$ = 30 K). 
}
\label{fig:FS} 
\end{figure}

Energy distribution curves (EDCs) along $\Gamma$--M in the normal state ($T$ = 30 K) of a NaFe$_{0.95}$Co$_{0.05}$As sample are plotted in
Fig.~\ref{fig:FS}(a). The corresponding color plot of second derivative $\partial^2I/\partial\omega^2$ as function of binding energy ($E_B$) and
in-plane momentum ($k$) along $\Gamma$--M is plotted in Fig.~\ref{fig:FS}(c).
We identify three hole-like bands around the $\Gamma$ point. Only one band, that we call $\alpha$,
clearly crosses $E_F$. 
According to LDA band calculations \cite{JPSJ.78.124712}, 
there exist three hole-like bands around $\Gamma$, two of them being degenerate at $\Gamma$ above the third one.
Following this scheme, we assume that a second band ($\alpha^\prime$) crosses $E_F$ and intersects with the $\alpha$ band at the $\Gamma$ point.
As illustrated in Fig.~\ref{fig:FS}(c), the intensity of the $\alpha^\prime$ band becomes much weaker when approaching toward $\Gamma$,
likely due to the matrix element effect.
In order to determine the top of the bands, we show the second derivative intensity plot along $\Gamma$--M at elevated temperature ($T$~=~100~K) in
Fig.~\ref{fig:FS}(b). The Fermi function was removed to unmask the band dispersion above $E_{F}$. The $\alpha$ band crosses $E_F$ at $k_F \sim
0.15\pi/a$ and reaches the top around 10-15 meV above $E_F$. A third band ($\beta$) that ``sinks" 10-15 meV below $E_{F}$ is observed
at 100 K [Fig.~\ref{fig:FS}(b)] and 30~K [Fig.~\ref{fig:FS}(c)].

In the vicinity of M, an electron-like band is clearly observed, as shown in the
second derivative plot along $\Gamma$--M in Fig.~\ref{fig:FS}(c). Along the perpendicular direction, as indicated by \#c3 in panel (d), the intensity
plot in Fig.~\ref{fig:FS}(e) shows two electron-like bands crossing $E_F$ with slightly different $k_F$ values from  Fig.~\ref{fig:FS}(c). The
observed bands are from two different electron-like bands \cite{JPSJ.78.124712}, refereed to as the $\gamma$, $\delta$ bands, in agreement with the
LDA calculations \cite{JPSJ.78.124712} and in an electron-doped ``122'' system \cite{Terashima2009}.

\begin{figure}[thb]
\vspace{-1em}
\begin{centering}
\includegraphics[width=3.4in]{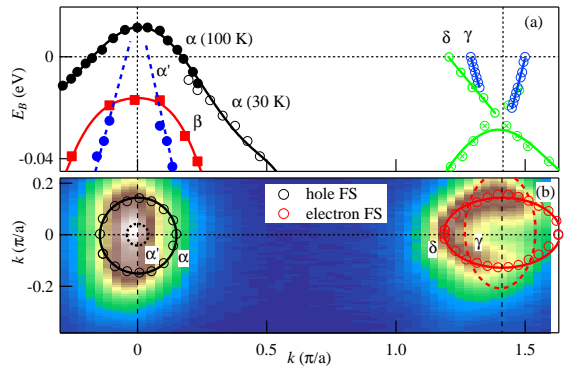} 
\end{centering}
\vspace{-1em}
\caption{(color online)
  (a) Extracted band dispersions along $\Gamma$--M in the normal state ($T$ =  30 K except for black solid circles [$T$ = 100 K]). The
  blue dashed lines are extensions of the $\alpha^\prime$ band, blue open circles are extracted dispersions through M and perpendicular to $\Gamma$--M.
  (b). Integrated intensity ($\pm$5 meV) around $E_F$ as function of ($k_x$, $k_y$). Open circles are extracted FSs. Solid lines are fitted FSs, while
  the black dashed circle is obtained from linear extension of the $\alpha^\prime$  band and the red dashed ellipse is a 90$^{o}$ rotation of the
  $\gamma$($\delta$) FS with respect to the M point.  
} 
\label{fig:Nesting}  
\end{figure} 
Figure \ref{fig:Nesting}(a) shows the extracted band dispersions from the data in Fig.~\ref{fig:FS}, with the suppressed portion of the
$\alpha^\prime$ band linearly extrapolated to $\Gamma$.
The estimated Fermi vector for the  $\alpha$ dash band is $k_F \leq 0.05\pi/a$. Figure \ref{fig:Nesting}(b) summarizes the FS topology of this sample: one
large circular hole-like FS ($\alpha$) and a possible much smaller one ($\alpha^\prime$) around $\Gamma$, and two elliptical electron-like FSs around
M ($\gamma(\delta)$). 
We estimate that the net enclosed FS area (the area of electron-like FS is positive, and the one for hole-like FS is negative, with $\pm 0.5^{\circ}$
angular resolution) is  about 4.3\%($\pm$1.2\%) of the
whole Brillouin zone, in agreement with the nominal bulk electron concentration of 5\% per Fe atom.
The $\alpha$ and $\gamma(\delta)$ bands can be connected fairly well by the Q = ($\pi$,$\pi$) AF  wavevector to allow inelastic interband
scattering. Even though AF fluctuations in this material were reported to be much weaker than in other pnictides \cite{Li2009p2147}, our  results are
qualitatively similar to that reported on other Fe-based superconductors, supporting inter-FS scattering over the near-nested FS as an important
ingredient to  superconductivity \cite{Chubukov2008lp,Mazin2008ta,Kuroki2008p3938}. 

\begin{figure}[thb]
\begin{centering}
\includegraphics[width=3.3in]{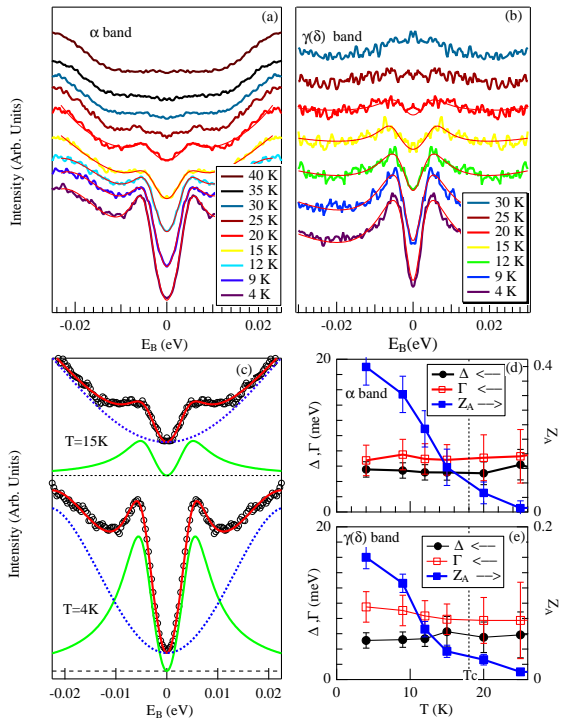} 
\end{centering}
\vspace{-1em}
\caption{
  (color online)
(a), (b) Temperature dependence of the $k_F$ spectra of the $\alpha$ and $\gamma(\delta)$ bands, respectively. The
$k$ location is indicated as p1 and p2 in Fig.~\ref{fig:FS}d, for $\alpha$ and $\gamma(\delta)$ bands respectively.
(c)  Fitting examples of symmetrized $\alpha$ band EDC spectra. Green lines correspond to $\Sigma(k,\omega)$ (see the text) whereas the dashed blue lines refer to a polynomial background. The fit results are given by the red lines. (d),(e) Temperature dependence of the quasi-particle properties from data in panels (a) and (b), respectively. $Z_A$ is the normalized coherent area given by $Z_A=\int A_{coh}(\omega) /A(\omega)d\omega$ in the [-20meV, 20meV] energy range.
}
\label{fig:Tdep} 
\vspace{-1em}
\end{figure}

Figure \ref{fig:Tdep} shows the temperature dependence of the SC coherent peaks and the SC gaps for the $\alpha$ and $\gamma(\delta)$ bands.
The symmetrized EDCs at $k_F$ measured at different temperatures are plotted in Fig.~\ref{fig:Tdep}(a) and \ref{fig:Tdep}(b).
Assuming particle-hole symmetry, this procedure allows us to approximately remove the effect of the Fermi-Dirac function. We define the SC gap value $\Delta(T)$ as half the distance between the two SC coherent peaks in the symmetrized EDCs. Sharp coherent peaks are seen at low temperature, with a peak-peak distance of $\sim$13 meV. 
With increasing temperature, the coherent peak intensity decreases steadily and becomes vanishingly small slightly above $T_c$.
In contrast to previously reported observation of a shoulder within SC gap in Ba$_{0.6}$K$_{0.4}$Fe$_2$As$_2$ \cite{Ding2008p3325}, whose origin
is still debated, we observe no shoulder here.
The gapped region between the coherent peaks also fills up and disappears at the same temperature as the peak vanishes. The observation of SC
coherence above $T_c$ is possibly due to the existence of a pseudogap with a crossover temperature ($T^\star$) about 5$\sim$10 K above $T_c$. An
alternative scenario is that the transition temperature at the surface, where the mobile Na concentration could be slightly different from the bulk,
is slightly higher than in the bulk, or the presence of SC fluctuations above $T_C$.

For a more quantitative analysis of the temperature evolution of the SC gap, we used the formula suggested by Norman \textit{et al.} describing the
self-energy $\Sigma(k,\omega)$ of quasi-particles in the SC state \cite{Norman1998p4522}:  $\Sigma(k,\omega)=-i\Gamma+\Delta^2
/[(\omega+i0^+)+\epsilon(k)]$, where $\Delta$ is the gap size and $\Gamma$ the single particle scattering rate simplified as $\omega$-independent.
The absence of shoulder within the SC gap allows us to perform reliable fits as compared to Ba$_{0.6}$K$_{0.4}$Fe$_2$As$_2$.
Assuming a polynomial background, this function fits the spectral lineshape remarkably well at different temperatures until the peak vanishes, as illustrated in Fig.~\ref{fig:Tdep}(c). The extracted gap sizes and coherent peak linewidths for the hole and electron bands are  shown in Fig.~\ref{fig:Tdep}(d) and \ref{fig:Tdep}(e), respectively. We observe that the gaps change little with increasing temperature and persist even above $T_c$. The linewidth ($\Gamma$, defined as half width at half maximum), which is proportional to the scattering rate, also changes little with temperature. On the other hand, the normalized coherent weight, defined as the integrated area  $Z_A = \int A_{coh}(k,\omega)d\omega / \int A(k,\omega) d\omega$ as following the same procedure as in the earlier work for the cuprates \cite{Ding2001p229,Feng2000p4551}, has a strong temperature dependence. As temperature increases, $Z_A$ decreases and approaches zero at $T_c$. 
This unusual temperature evolution is different from the ``gap-closing" behavior of a conventional BCS superconductor, and is similar to what has been observed in the underdoped cuprates \cite{Norman1998p4522}, suggesting unconventional superconductivity in this pnictide superconductor.

\begin{figure}[htb]
\begin{centering} 
\includegraphics[width=3.4in]{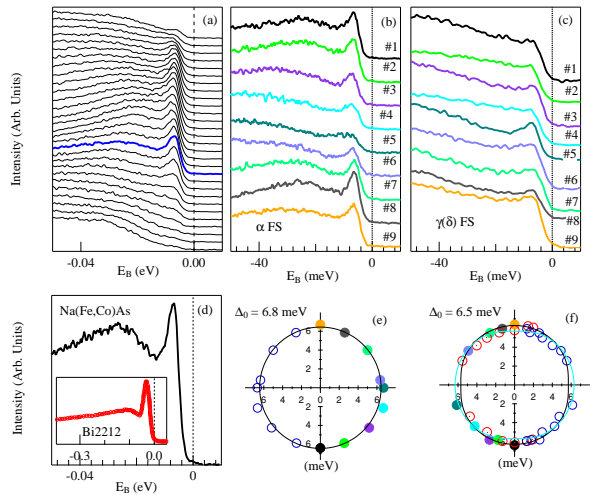} 
\end{centering}
\vspace{-2em}
\caption{(color online)
  (a) EDCs of the $\alpha$ band along the (0,0)--(-$\pi,\pi$) direction in the SC state ($T$ = 4 K). 
  (b),(c) EDCs along the $\alpha$ and $\gamma(\delta)$ FSs, respectively.
 (d) Typical EDC at $k_F$ for NaFe$_{0.95}$Co$_{0.05}$As $\alpha$ band at $T$ = 8 K.
   The red curve is an anti-nodal EDC of optimally-doped Ba$_2$Sr$_2$CaCu$_2$O$_{8+x}$ at $T\ll T_c$
  \cite{Ding2001p229}, showing the well known peak-dip-hump structure. 
  (e) Extracted gap size from EDCs along the FSs, plotted in polar coordinates with respect to their FS center. The color of the dots refer to the
  color of the EDCs in (b). Solid symbols correspond to additional extracted values while
  blue circles are obtained by symmetry operations. Black lines are $\Delta(s_{\pm})=\Delta_0 \vert \cos k_x \cos k_y \vert$ fits, with $\Delta_0$ = 6.8
  meV. 
  (f) Same as (e) but for the $\gamma(\delta)$ band. The color scale refers to (c). Black and cyan lines are $\Delta(s_{\pm})$ fits with $\Delta_0=6.5$ meV.
}
\label{fig:Kdep} 
\end{figure}
 
The symmetry of the SC order parameter is arguably the most important information in understanding the SC mechanism of a superconductor. Thus we have performed high-resolution measurements of the $k$-dependence of the SC gap for the $\alpha$ and $\gamma(\delta)$ bands. One example is given in Fig.~\ref{fig:Kdep}(a), which shows the EDCs across $k_F$ of the $\alpha$ band in the  SC state ($T$ = 4 K). The band disperses toward $E_F$ and starts to bend back at $k_F$, with the minimum gap situating at $\sim$ 6--7 meV below $E_F$. 
Similar behavior was observed for the electron-like $\gamma(\delta)$ bands. We also observed a kink in the band dispersion at $\omega \sim$ 11 meV,
which could be due to a coupling between electrons and a collective mode \cite{Richard2009p3322}. EDCs at various $k_F$ along the $\alpha$ and
$\gamma(\delta)$ FSs are plotted in Figs.~\ref{fig:Kdep}(b)~and~(c), respectively. Here we compare the spectral lineshape of NaFe$_{0.95}$Co$_{0.5}$As
and the  cuprate Ba$_2$Sr$_2$CaCu$_2$O$_{8+x}$. The spectrum of the pnictide ($\alpha$ band) shown in Fig.~\ref{fig:Kdep}(d) has a  sharp
quasiparticle peak at the gap edge followed by a broad peak which is due to an extra band below the $\alpha$ band.  The cuprate has a similar
peak-dip-hump structure near ($\pi$, 0), as shown by the red curve in the inset (obtained from Ref.~\cite{Ding2001p229}), although it has a more
complicated origin. We note that the coherent peak width of the pnictide ($\sim$ 6--7 meV) is much smaller than the one of the cuprate ($\sim$ 20
meV), suggesting a smaller gap inhomogeneity in this pnictide as compared to the cuprate~\cite{Pan2001p5619}.

The coherent peak of the $\alpha$ band is always located about 6.5--7 meV below $E_{F}$, with very small variation (less than $\pm 0.5$ meV) around the
FS, giving a $2\Delta/k_{B}T_{c} \sim  8-9$ ratio. A similar value is obtained for the $\gamma(\delta)$ FSs, indicating that
NaFe$_{0.95}$Co$_{0.05}$As is in the strong coupling regime. The extracted SC gap sizes at low temperature [$\Delta(0)$] for the $\alpha$ and
$\gamma(\delta)$ bands are plotted in Fig.~\ref{fig:Kdep}(e) and (f), respectively, as a function of polar angle with respect to their FS
center. There is no gap node or large variation in the gap size for all the FSs. This excludes the possibility of symmetries with nodes such as
$d$-wave. The natural candidates are the $s$-wave or $s_{\pm}$-wave symmetries. In  Fig.~\ref{fig:Kdep}(e) and \ref{fig:Kdep}(f), the solid lines
indicate fits to the simple $s_{\pm}$-wave gap function of $\Delta_0 \vert \cos k_{x}\cos k_{y} \vert$ \cite{Seo2008p5623}, with $\Delta_0$ = 6.8 meV
and 6.5 meV for the hole and electron FSs, respectively. Clearly, the $k$-dependence of the SC gaps in this 111 system is in good agreement with a
$s_{\pm}$-wave, similarly to the 122 system Ba$_{0.6}$K$_{0.4}$Fe$_2$As$_2$ \cite{Ding2008p3325, Nakayama2009p3324}. 
 
We comment on a recent ARPES study on SC LiFeAs \cite{Borisenko2010p4526}. Using the leading shift across $T_c$ to define the SC gap size, it claimed
a much smaller $2\Delta/k_{B}T_{c}\sim3.5$ ratio and suggested phonon-mediated BCS superconductivity in LiFeAs. LiFeAs has a smaller As-Fe-As angle
and is nonmagnetic, which might cause the striking difference between LiFeAs and the other strongly coupled Fe-based superconductors, which now
includes the NaFe$_{0.95}$Co$_{0.05}$As compound with the same crystal structure as LiFeAs. Another possibility is the higher mobility of the Li atoms
on the surface as compared to Na, which may lead to a different stoichiometry at the surface. In addition, the leading edge shift may likely
underestimate the SC gap size, which would in turn yield a smaller $2\Delta/k_{B}T_{c}$ ratio. Our observation of the large $2\Delta/k_B T_c$ ratio
and the unusual temperature dependence of the SC gap in NaFe$_{0.95}$Co$_{0.05}$As strongly suggest unconventional superconductivity in this material.

We acknowledge useful discussions with Ziqiang Wang, Jiangping Hu and Weiqiang Yu. The experiment was
supported by the grants from NSF, MOST and CAS of China, and JSPS, JST-TRIP, JST-CREST, MEXT of Japan.
The Synchrotron Radiation Center, WI is supported by NSF of US. 

\bibliographystyle{unsrtnat} 
\bibliography{reference}
\end{document}